\documentclass[prd,preprint,nofootinbib]{revtex4-1}

\newcount\rem \rem=1
\ifnum \rem=0 
\newcommand{\ygn}[1]{}
\newcommand{\adn}[1]{}
\newcommand{\mgn}[1]{}
\newcommand{\ssn}[1]{}
\else
\newcommand{\ygn}[1]{{\bf \color{red} [YG:~#1]}}
\newcommand{\adn}[1]{{\bf \color{blue} [AD:~#1]}}
\newcommand{\mgn}[1]{{\bf \color{orange} [MG:~#1]}}
\newcommand{\ssn}[1]{{\bf \color{magenta} [S.S.:~#1]}}
\fi

\usepackage{amssymb,amsmath}
\usepackage{color}
\usepackage[toc,page]{appendix}
\usepackage{graphicx}
\usepackage{slashed}
\usepackage{tikz,subfig}
\usepackage{hyperref}

\newcommand{\nn}{\nonumber}

\renewcommand{\Im}{{\cal I}m}
\renewcommand{\Re}{{\cal R}e}

\newcommand{\D}{\mathrm{d}}
\newcommand{\half}{\frac{1}{2}}

\newcommand{\ket}[1]{|{#1} \rangle}
\newcommand{\bra}[1]{\langle {#1}|}

\newcommand{\mel}[3]{\langle{#2}|{#1}|{#3}\rangle}

\newcommand{\ks}{\ket{K_S}}
\newcommand{\kl}{\ket{K_L}}
\newcommand{\kb}{\ket{\overline{K}{}^0}}
\newcommand{\kbar}{\overline{K}{}^0}
\newcommand{\ko}{\ket{K^0}}
\newcommand{\kzero}{K^0}

\newcommand{\xtg}{\Delta m t}

\def\beq{\begin{eqnarray}}
\def\eeq{\end{eqnarray}}
\def\no{\nonumber}

\newcounter{mycount}
\newcommand{\pauseen}{\setcounter{mycount}{\value{enumi}}\end{enumerate}}
\newcommand{\resumeen}{\begin{enumerate}\setcounter{enumi}{\value{mycount}}}

\begin{document}
\title{$K\to \mu^{+} \mu^{-}$ as a clean probe of short-distance physics}
\author{Avital Dery}
\email{avital.dery@cornell.edu}
\affiliation{Department of Physics, LEPP, Cornell University, Ithaca, NY 14853, USA}
\author{Mitrajyoti Ghosh}
\email{mg2338@cornell.edu}
\affiliation{Department of Physics, LEPP, Cornell University, Ithaca, NY 14853, USA}
\author{Yuval Grossman}
\email{yg73@cornell.edu}
\affiliation{Department of Physics, LEPP, Cornell University, Ithaca, NY 14853, USA}
\author{Stefan Schacht}
\email{stefan.schacht@manchester.ac.uk}
\affiliation{Department of Physics and Astronomy, University of Manchester, Manchester, M13 9PL, United Kingdom}

\preprint{MAN/HEP/2021/003}

\begin{abstract}

The $K\to\mu^+\mu^-$ decay is often considered to be
uninformative of fundamental theory parameters since the decay is
polluted by long-distance hadronic effects. We demonstrate that, using
very mild assumptions and utilizing time-dependent interference effects, 
${\cal B}(K_S\to\mu^+\mu^-)_{\ell=0}$ can be experimentally determined
without the need to separate the $\ell=0$ and $\ell=1$ final states.
This quantity is very clean
theoretically and can be used to
test the Standard Model. In particular, it can be used to 
extract the CKM matrix
element combination 
$\left|V_{ts}V_{td}\sin(\beta+\beta_s)\right|\approx |A^2 \lambda^5 \bar \eta|$
with hadronic uncertainties below $1\%$.
\end{abstract}

\maketitle

\section{Introduction}\label{sec:intro}

Rare flavor changing neutral current (FCNC) kaon decays \cite{Inami:1980fz, Hagelin:1989wt, Dib:1991mx, Buchalla:1993wq, DAmbrosio:1996lam, Pich:1996sx, Buras:1997fb} provide a unique way to probe the flavor sector of
the Standard Model (SM) and, in particular, CP-violating effects. The program to measure
the decay rates of $K^+ \to \pi^+ \nu \bar\nu$~\cite{CortinaGil:2020vlo} and $K_L \to \pi^0 \nu \bar \nu$~\cite{Ahn:2020opg} is
aimed at determining the CKM parameters with very high
theoretical precision. In particular, the $K_L \to \pi^0 \nu \bar \nu$
decay rate can be used to extract~\cite{Buras:2015qea,Buchalla:1998ba,Mescia:2007kn}
\beq\label{eq:ImVtsVtd}
\left|V_{ts}V_{td}\,\sin(\beta+\beta_s)\right| \approx |A^2 \lambda^5 \bar\eta|\,,
\eeq
where $A$, $\lambda$, and $\bar\eta$ are the Wolfenstein parameters and $\beta+\beta_s$ is one of the angles in the $ds$ unitarity
triangle such that~\cite{Lebed:1996ay}
\beq
\beta= \arg\left(-\frac{V_{cd}V_{cb}^*}{V_{td} V_{tb}^*}\right),
  \qquad
\beta_s = \arg\left(-\frac{V_{ts} V_{tb}^*}{V_{cs} V_{cb}^*}\right), \qquad
\beta+\beta_s-\pi=\arg\left(-\frac{V_{ts} V_{td}^*}{V_{cs} V_{cd}^*}\right). 
\eeq

Experimentally, working with decays that involve charged leptons is
much simpler than the above-mentioned neutrino
modes. 
Nonetheless, the focus of the current kaon program is on the
neutrino final states, primarily because decays to charged leptons are believed not to be theoretically clean.
There are so-called long-distance effects that introduce hadronic uncertainties, making extractions of clean 
theory parameters challenging. 

In this paper, we show that we can get very clean theoretical
information from decays of kaons into charged leptons. This can be done
only for the neutral kaons, by exploiting the
interference effects between $K_S$ and $K_L$. 
We focus on $K \to \mu^+ \mu^-$, for which the relevant CKM observable
is that of Eq.~(\ref{eq:ImVtsVtd}). 
The theoretical precision in this case is superb, 
with hadronic uncertainties below the $1\%$
level.

The importance of the interference terms in 
$K \to \mu^+ \mu^-$ was emphasized in
Ref.~\cite{DAmbrosio:2017klp}. In this paper, we generalize their
results and demonstrate that one can get a very clean determination of
the parameter combination in Eq.~(\ref{eq:ImVtsVtd}) by studying the interference terms.

Before we get into the details, below we explain the main idea. We first recall the situation with $K_L \to \pi^0 \nu
\bar\nu$. The reason that this decay mode is theoretically clean is that it is to
a very good approximation pure CP-violating. As such, it is all
calculable using perturbation theory and we do not have to worry
about non-calculable long-distance effects, as they are to a very good
approximation CP conserving.

The issue with $K \to \mu^+ \mu^-$ is that the final state is a
mixture of $\ell=0$ and $\ell=1$ partial wave configurations. Thus, both $K_S$ and $K_L$ decays
are not pure CP-violating, and both decays have non-calculable long
distance effects.
Yet, if we could experimentally distinguish between the $\ell=0$ and $\ell=1$ final
states, the situation would be similar to  $K_L \to \pi^0 \nu
\bar\nu$, as we could separate the CP-violating part that we can
calculate. In particular, the $\ell=0$ amplitude has
significant CP violation effects in the SM, and the decay mode $K_S
\to (\mu^+ \mu^-)_{\ell=0}$ is very clean theoretically.
What we show in this work is that under some mild assumptions
we can extract the rate, that is,
${\cal B}(K_S\to (\mu^+ \mu^-)_{\ell=0})$
without separating the $\ell=0$ and $\ell=1$
final states. This can be done by isolating the interference terms.

Leptonic kaon decays have been studied for a long time~\cite{Martin:1970ai, Pais:1969bi, Sehgal:1969zk, Christ:1971hr,Sehgal:1972db,Dass:1972cq,  Gaillard:1974hs,Gaillard:1975ds,Buras:1981cr, Bergstrom:1983zf, Geng:1989ah,Gaillard:1975ds,Bogomolny:1975nz,Lee:1972px, Voloshin:1976hp, Shrock:1979eb, Herczeg:1982sf, Stern:1973xs}. 
Rare kaon decays have a lot of potential for the discovery of physics
beyond the SM~\cite{Grossman:1997sk,
  Aebischer:2020mkv,Mandal:2019gff,Endo:2017ums,Bobeth:2017ecx,
  Chobanova:2017rkj, Bobeth:2017xry, Endo:2016tnu, Tanimoto:2016yfy,
  Buras:2015jaq,Buras:2015yca,Blanke:2008yr,  Mescia:2006jd,
  Deshpande:2004xc, Crivellin:2016vjc, Kitahara:2019lws,
  Ziegler:2020ize, He:2020jly, Gori:2020xvq}.
Also on the experimental side a lot of advances took place in the quest for rare kaon decays~\cite{Aaij:2020sbt, Ambrosino:2019qvz, Aaij:2017tia, AmelinoCamelia:2010me, Abouzaid:2007cm, Ahn:2020opg, Lollini:2021lue,CortinaGil:2020vlo, Cerri:2018ypt}.

The SM predictions for $K\rightarrow \mu^+\mu^-$~\cite{Ecker:1991ru, Isidori:2003ts, DAmbrosio:2017klp, Gorbahn:2006bm} and the
corresponding long-distance contributions~\cite{Ecker:1991ru, Colangelo:2016ruc, GomezDumm:1998gw, Knecht:1999gb} have been studied in 
great detail. The same goes for $K_S\rightarrow \gamma\gamma$ and $K_S\rightarrow \gamma l^+l^-$~\cite{Colangelo:2016ruc}
as well as kaon decays into four leptons~\cite{DAmbrosio:2013qmd}. See also the reviews Refs.~\cite{Cirigliano:2011ny, Buras:2004uu}.

\section{Notation and formalism}\label{sec:notation}

We use the following standard notation~\cite{Zyla:2020zbs}, where the two neutral kaon mass eigenstates, $\ks$ and $\kl$, are linear combinations of the flavor eigenstates:
\beq 
\ket{K_{S}} = p \ko + q \kb,\qquad \ket{K_{L}} = p \ko - q \kb.
\eeq
The mass and width averages and differences are denoted by
\beq 
m &=& \frac{m_L+m_S}{2}, \qquad \Gamma = \frac{\Gamma_L+\Gamma_S}{2}, \\ \no
\Delta m &=& m_L-m_S, \qquad \Delta \Gamma = \Gamma_L-\Gamma_S. 
\eeq
We define the decay amplitudes of $\ko$ and $\kb$ to a final state $f$,
\beq 
A_f = \mel{{\cal H}}{f}{K^0},  \qquad \overline A_{f} = \mel{{\cal
    H}}{f}{\overline{K}{}^0}, \label{eq:Af-Afbar} 
\eeq
and the parameter $\lambda_f$,
\beq \lambda_f \equiv \frac{q}{p} \frac{\overline{A}_f}{A_f}\,. \eeq
We use an arbitrary normalization, such that $A_f$ and
$\overline{A}_f$ have the same normalization. 

An amplitude is called
{\it relatively real} if $\Im \lambda_f =0$ and {\it relatively
  imaginary} if $\Re \lambda_f =0$. Any amplitude can be written as a
sum of a relatively real and a relatively imaginary part.

In any neutral meson system, the quantities $A_f$, $\overline{A}_f$,
and $q/p$ depend on the phase convention. However, $|A_f|$,
$|\overline{A}_f|$, $|q/p|$, and $\lambda_f$ are phase convention
independent and are hence physical. 

Consider a beam of neutral kaons.
The time dependent decay rate as a function 
of proper time  is given by~\cite{Zyla:2020zbs} 
\beq \label{eq:time-dep}
\left(\frac{\D \Gamma}{\D t} \right) = 
{\cal N}_f f(t),
\eeq
where ${\cal N}_f$ is a time-independent normalization factor and 
the function $f(t)$ is given as a sum of four functions
\beq \label{eq:Cdef}
f(t) = C_L e^{-\Gamma_L t}+
C_S \,e^{-\Gamma_S t} + 2 \left[C_{sin} \sin (\xtg) + C_{cos} \cos
(\xtg)\right]e^{-\Gamma t}. 
\eeq
The form of Eq.~(\ref{eq:Cdef}) is valid for any neutral kaon beam 
(that is, not only for a pure state) and also for a sum over several
final states.
We refer to the set of coefficients, $\{C_L,C_S,C_{sin},C_{cos}\}$, as
the {\it experimental parameters}.
Note that $C_L$ is the coefficient of the $K_L$ decay term, 
$C_S$ of the $K_S$ decay term, while $C_{\sin}$ and $C_{\cos}$ come with the
interference terms between $K_L$ and $K_S$. 
For convenience we also define
\beq\label{eq:defCInt}
C_{Int.}^2 = C_{cos}^2 + C_{sin}^2\,. 
\eeq

The $C$ coefficients implicitly depend on the composition of the beam
and on the relevant final states. The dependence on the final states enters via 
the parameters 
\begin{align}
\{|A_f|\,, \qquad 
|\overline{A}_f|\,, \qquad
\vert q/p\vert\,, \qquad 
\arg(\lambda_f)\}.
\end{align}
We denote these as the {\it theory parameters}.

For an initial $\ko$ and $\kb$ beam, respectively, and a single final state, $f$, the coefficients are explicitly given by~\cite{Zyla:2020zbs}
\begin{align}
C_L^{K^0} &= \frac{1}{2} |A_f|^2\left(1+|\lambda_f|^2-2\Re\lambda_f\right)\,, & 
C_L^{\overline K{}^0} &=
        \frac{1}{2}|\overline A{}_f|^2\left(1+|\lambda_f|^{-2}-2\Re\lambda_f^{-1}\right), \nonumber \\[7pt] \nonumber
	C_S^{K^0} &= \frac{1}{2}|A_f|^2\left(1+|\lambda_f|^2+2\Re\lambda_f\right)\,, & 
	C_S^{\overline K{}^0} &= \frac{1}{2}|\overline A{}_f|^2\left(1+|\lambda_f|^{-2}+2\Re\lambda_f^{-1}\right), \\[7pt] \nonumber
	C_{\sin}^{K^0} &= -|A_f|^2\Im\lambda_f\,, & 
	C_{\sin}^{\overline K{}^0} &= -|\overline A{}_f|^2\Im\lambda_f^{-1}, \\[7pt] 
	C_{\cos}^{K^0} &= \frac{1}{2} |A_f|^2\left(1-|\lambda_f|^2\right)\,, & 
	C_{\cos}^{\overline K{}^0} &= \frac{1}{2} |\overline A{}_f|^2\left(1-|\lambda_f|^{-2}\right). \label{eq:coefficients}
\end{align}

In the following we focus on decays into CP-eigenstate final
states. For a given final state, $f$, we define $\eta_f=1$ if it is CP-even and
$\eta_f=-1$ if it is CP-odd.
We define the CP-even and CP-odd amplitudes
\beq
A_f^\text{CP-even} \equiv \frac{1}{\sqrt{2}} \,A_f\,(1+\eta_f \lambda_f)\,, \qquad
A_f^\text{CP-odd}\equiv \frac{1}{\sqrt{2}} \,A_f\,(1-\eta_f \lambda_f)\,.
\label{eq:defacpeo}
\eeq

We make several assumptions and approximations as we go on. Our first
approximation is
\begin{itemize}
\item [$(i)$]
{\it CP violation (CPV) in mixing is negligible.}  \\
Although our main interest is CP violating physics, CPV in mixing is
sub-dominant in the effects we consider. We therefore neglect it
throughout the paper and work in the limit 
\beq
	\Big\lvert \frac{q}{p}\Big\rvert = 1.
\eeq
\end{itemize}
This approximation is known to work to order $\epsilon_K \sim
10^{-3}$ which we neglect from this point on.

Under the above assumption, 
the full set of decay-mode-specific independent physical parameters can be taken to be
\beq\label{eq:setlambda}
	\{|A_f|,\qquad |\overline{A}_f|,\qquad \arg\left(\lambda_f\right)\}.
\eeq
Furthermore, in the limit of no CPV in mixing, the CP
amplitudes of Eq.~(\ref{eq:defacpeo}) correspond to the amplitudes for the decays of $K_S$ and $K_L$. For
example, for $f=\pi^+\pi^-$, $\eta_f=1$ and to a very good approximation $\lambda_f = 1$
and thus $A_{\pi^+\pi^-}^\text{CP-odd} =0$. In the case of $K \to \pi \nu \bar \nu$, $\eta_f=1$ and $\lambda_f$ is to a very good
approximation a pure phase, so that the amplitude for $K_L \to \pi \nu \bar
\nu$ gives sensitivity to the phase $\arg(\lambda_f)$~\cite{Grossman:1997sk}. 

In the following it will be useful to replace the set of independent physical parameters of
Eq.~(\ref{eq:setlambda}) 
with the equivalent set of physical parameters:
\beq
\{|A^\text{CP-even}_{f}|,\qquad |A^\text{CP-odd}_{f}|,\qquad \arg\left({A_{f}^\text{CP-even}}^*A^\text{CP-odd}_{f}\right)\}.
\eeq
In particular, the time dependence for a beam of initial $\ko$ into a 
CP-even final state is given by the coefficients
\beq\label{eq:K0tofCs}
C_L^{K^0} = |A_f^\text{CP-odd}|^2, &&\qquad 
C_S^{K^0} = |A_f^\text{CP-even}|^2,  \nn\\ 
C_{cos}^{K^0} = \Re(A_f^\text{CP-odd*}A_f^\text{CP-even}),&&
\qquad
C_{sin}^{K^0} = -\Im(A_f^\text{CP-odd*}A_f^\text{CP-even}),\label{eq:coefficients-CP-even}
\eeq 
For a CP-odd final state it is given by 
\beq
C_L^{K^0} = |A_f^\text{CP-even}|^2, &&\qquad
C_S^{K^0} = |A_f^\text{CP-odd}|^2,  \nn\\ 
C_{cos}^{K^0} = \Re(A_f^\text{CP-odd*}A_f^\text{CP-even}),&&
\qquad
C_{sin}^{K^0} = \Im(A_f^\text{CP-odd*}A_f^\text{CP-even}).
\label{eq:coefficients-CP-odd}
\eeq
For an initial $\kb$ state the result is obtained by multiplying
$C_{cos}$ and $C_{sin}$ by $-1$ in  
Eqs.~(\ref{eq:K0tofCs}) and (\ref{eq:coefficients-CP-odd}).

We also define
\beq
\varphi_f=\arg(A_f^\text{CP-odd*}A_f^\text{CP-even}),
\eeq
such that we can write for a CP-even final state
\beq
C_{cos}^{K^0}=  |A_f^\text{CP-odd*}\,A_f^\text{CP-even}|  \cos\varphi_f\,, \qquad
C_{sin}^{K^0} = -|A_f^\text{CP-odd*}\,A_f^\text{CP-even}|  \sin\varphi_f\,.
\eeq
For a CP-odd final state we have analogously 
\begin{align}
C_{cos}^{K^0} &= |A_f^\text{CP-odd*}\,A_f^\text{CP-even}|  \cos\varphi_f\,, & 
C_{sin}^{K^0} &= |A_f^\text{CP-odd*}\,A_f^\text{CP-even}|  \sin\varphi_f\,.
\end{align}

\section{The $K\to\mu^+\mu^-$ decay} \label{sec:kmumu}
In the decay of a neutral kaon into a pair of muons, there are two
orthogonal final states that are allowed by conservation of angular
momentum --- muons with a symmetric wave function ($\ell =0$) and muons
with an anti-symmetric wave function ($\ell =1$). Note that since the
leptons are fermions, the state with $\ell = 0$ has negative parity
and so it is CP odd, and the state with $\ell = 1$ is CP even.
The four relevant amplitudes can be written in terms of the CP amplitudes of Eq.~(\ref{eq:defacpeo}) as
\beq\label{eq:acpoddeven}
A_\ell^\text{CP-even} =  \frac{1}{\sqrt{2}} \,A_\ell\,\Big(1- (-1)^{\ell} \lambda_\ell\Big), \qquad
A_\ell^\text{CP-odd} =  \frac{1}{\sqrt{2}} \,A_\ell\,\Big(1+(-1)^{\ell} \lambda_\ell\Big),
\eeq 
with $\ell=0,1$.
Note that we keep the normalization arbitrary, but if we want to maintain the same normalization for both $A_0$ and $A_1$ then we require a relative 
phase space factor between them, $\beta^2_{\mu}$, with
\begin{align}
\beta_\mu \equiv \left(1-\frac{4m_\mu^2}{m_K^2}\right)^\half\,,
\end{align}
see for details Appendix~\ref{sec:appSM}.

Note that under the approximation $|q/p|=1$, Eq.~\eqref{eq:acpoddeven} allows us to write the CP-even and -odd amplitudes as amplitudes for the decays of the mass eigenstates $\ks$ and $\kl$:
\beq
A_0^\text{CP-odd} &=& A(K_S\to \mu^+\mu^-)_{\ell = 0}\,, \nonumber\\
A_0^\text{CP-even} &=& A(K_L \to \mu^+\mu^-)_{\ell = 0}\,, \nonumber\\
A_1^\text{CP-odd} &=& A(K_L\to \mu^+\mu^-)_{\ell = 1}\,, \nonumber\\
A_1^\text{CP-even} &=& A(K_S \to \mu^+\mu^-)_{\ell = 1}.
\eeq

When measuring the total time dependent decay rate for
$K\to\mu^+\mu^-$, the two di-muon configurations, $\ell=0,1$ add
incoherently. The form of the function $f(t)$ defined in
Eq.~(\ref{eq:Cdef}), is unchanged. 
Theoretically,  each of the $C$'s is given by an implicit sum over the
relevant amplitude expressions for different $\ell$'s.
Thus we have two sets of decay-mode-specific physical theory parameters,
\beq
\{|A^\text{CP-even}_{\ell}|,\qquad |A^\text{CP-odd}_{\ell}|,\qquad
\varphi_\ell \equiv\arg\left(A_{\ell}^\text{CP-odd*}A^\text{CP-even}_{\ell}\right)\}, 
\eeq
with $\ell=0,1$,
bringing us to a total of six unknown physical parameters.

It is well known that the decay $K\to\mu^+\mu^-$ receives long-distance and short-distance contributions~\cite{Littenberg:2002um,Greynat:2003ja,DAmbrosio:1997eof,Isidori:2003ts}. 
The long-distance contribution is dominated by diagrams with two intermediate on-shell photons, 
while the short-distance contribution is defined as originating from
the weak effective Hamiltonian. The distinction between long-distance and short-distance physics is somewhat ambiguous. It is
clear that the short-distance physics is to a good approximation dispersive
(real), since it is dominated by heavy particles in the 
loops. However, long-distance
diagrams contribute both to the absorptive (imaginary) amplitude and,
when taken off-shell, also to the dispersive amplitude. 

In the following we make one extra simplifying
assumption, which results in reducing the number of unknown parameters for
$K\to\mu^+\mu^-$. We consider only models where
\begin{itemize}
\item[$(ii)$]
{\it The only source of CP violation is in the $\ell=0$ amplitude.} \\
What we mean by this assumption is that only the $\ell=0$ amplitude
has $\Im(\lambda_\ell)\ne~0$.
\end{itemize}
As we discuss in Section~\ref{sec:sdphy-SM} and in
Appendix~\ref{sec:l0-only}, this assumption is fulfilled to a very good approximation within the SM and in any model in which the leading leptonic operator is vectorial. 

We can then draw an important conclusion from the above
assumption:
\beq
A^\text{CP-odd}_1=0.
\eeq
This implies that
the number of unknown parameters is reduced by two, 
leaving a single parameter, $|A_1^\text{CP-even}|$, for the $\ell=1$ final state.
Thus, we are left with the following list of four unknown physical parameters,
\beq
|A_0^\text{CP-odd}|,\qquad|A_0^\text{CP-even}|,\qquad|A_1^\text{CP-even}|,
\qquad \arg(A_0^\text{CP-odd*}A_0^\text{CP-even}).
\eeq
In the rest of the paper we demonstrate how 
it is possible to extract these parameters, and
specifically $|A^\text{CP-odd}_{0}| = A(K_S\to \mu^+\mu^-)_{\ell = 0}$, which, as we explain below, is a clean probe of the SM.

\section{Extracting ${\cal B}(K_S \to \mu^+\mu^-)_{\ell=0}$}\label{sec:sdphy}

As portrayed in Eq.~(\ref{eq:Cdef}), the time-dependent decay rate
for an arbitrary neutral kaon initial state is given in general by the
sum of four independent functions of time that depend on the
experimentally extracted parameters
\beq
\{C_L, \qquad C_S, \qquad C_{cos}, \qquad C_{sin}\}.
\eeq 
Within our assumptions, these coefficients depend on the following
four theory parameters
\beq\label{eq:physparam}
\{|A_0^\text{CP-odd}|,\qquad
|A_0^\text{CP-even}|,\qquad|A_1^\text{CP-even}|, \qquad \varphi_0 \equiv \arg(A_0^\text{CP-odd*}A_0^\text{CP-even})\}.
\eeq

We consider a case of a beam that at $t=0$ was a pure $K^0$ beam (that is,
no $\overline K{}^0$). Using Eq.~(\ref{eq:coefficients}) we obtain
that the result for this case is given by
\beq \label{eq:K-Cs}
	C_L &=&  |A_0^\text{CP-even}|^2, \\ \nonumber
C_S &=& |A_0^\text{CP-odd}|^2 + \beta_{\mu}^2 |A_1^\text{CP-even}|^2 , \\ \nonumber
C_{cos} &=& \Re(A_0^\text{CP-odd*}A_0^\text{CP-even})=
        |A_0^\text{CP-odd*}A_0^\text{CP-even}|\,\cos\varphi_0,  \\ \nonumber
C_{sin} &=& \Im(A_0^\text{CP-odd*}A_0^\text{CP-even})=|A_0^\text{CP-odd*}A_0^\text{CP-even}|
      \sin\varphi_0.
\eeq 
We see that the four experimental parameters can be used to extract the four
theory parameters.
In particular, we find 
\beq\label{eq:magic}
|A_0^\text{CP-odd}|^2 = \frac{C_{cos}^2 + C_{sin}^2}{C_L} = 
\frac{C_{Int.}^2}{C_L}\,,
\eeq
where $C_{Int.}^2 = C_{cos}^2 + C_{sin}^2$ was defined in Eq.~(\ref{eq:defCInt}).
Having the magnitude of the amplitude
we can deduce the branching ratio in terms of other observables,
\beq \label{eq:magicBR}
{\cal B}(K_S \to \mu^+\mu^-)_{\ell=0} = 
{\cal B}(K_L \to \mu^+\mu^-) \times {\tau_S \over \tau_L} \times
\left({C_{int}\over C_L}\right)^2\,.
\eeq
Eq.~(\ref{eq:magicBR}) is our main result. 
It demonstrates that we can extract ${\cal B}(K_S \to
\mu^+\mu^-)_{\ell=0}$ from the experimental time dependent decay rate.

A few comments are in order regarding Eq.~(\ref{eq:magicBR}):
\begin{enumerate}
\item 
Our ability to extract ${\cal B}(K_S \to
\mu^+\mu^-)_{\ell=0}$ comes from the interference
terms. It cannot be extracted from pure $K_L$ or $K_S$ terms.  
\item
A measurement of the interference terms additionally amounts to a measurement of the phase $\varphi_0$, which is not calculable from short-distance physics.
\item
In order to extract  ${\cal B}(K_S \to
\mu^+\mu^-)_{\ell=0}$ we need only three of the four experimental parameters. The fourth parameter, $C_S$, can then be used to extract
$|A_1^\text{CP-even}|$, or equivalently ${\cal B}(K_S \to
\mu^+\mu^-)_{\ell=1}$. Yet, this is not our main interest, as $|A_1^\text{CP-even}|$
is not calculable from short-distance physics.
\item
For a pure $\overline K{}^0$ beam, $C_S$ and $C_L$ in Eq.~(\ref{eq:K-Cs})
are unchanged while $C_{cos}$ and $C_{sin}$ pick up a minus sign, and
Eq.~(\ref{eq:magicBR}) is unchanged.
\end{enumerate} 

While we have only discussed a pure $K^0$ beam in this section, 
as long as we have sensitivity to the interference terms,
it is possible to determine $|A_0^\text{CP-odd}|$.
In particular, 
as long as the kaon decays in vacuum, one can write the branching ratio ${\cal B}(K_S \to \mu^+\mu^-)_{\ell=0}$ in terms of ${\cal B}(K_L \to \mu^+\mu^-)$ in the following way:
\beq \label{eq:magicBRgen}
{\cal B}(K_S \to \mu^+\mu^-)_{\ell=0} = {\cal D}_F \times
{\cal B}(K_L \to \mu^+\mu^-) \times {\tau_S \over \tau_L} \times \left({C_{int} \over C_L}\right)^2.
\eeq
where
${\cal D}_F$ is a dilution factor that takes into account the
particular composition of the kaon beam.
We discuss two cases, that of a mixed beam, and of a $K_L$ beam
with regeneration, in Appendix~\ref{sec:tDandr}.

\section{Calculating ${\cal B}(K_S \to \mu^+\mu^-)_{\ell=0}$}\label{sec:sdphy-SM}

We move to discuss the theoretical calculation of 
${\cal B}(K_S \to \mu^+\mu^-)_{\ell=0}$. 

\subsection{General calculation}
We define 
\beq
A_\ell=A_\ell^{SD} + A_\ell^{LD}.
\eeq
The short-distance (SD) amplitude, $A_\ell^{SD}$, is the one that can be calculated
perturbatively from
the effective Hamiltonian of any model. Note that at leading order in the perturbative
calculation it carries no strong phase. By definition, the long-distance (LD) amplitude, $A_\ell^{LD}$, is 
the part that is not captured by that calculation. In general, it
carries a strong phase.
We further define
\beq
\lambda_\ell^{SD} = {q \over p} {\overline{A}_\ell^{SD} \over A_\ell^{SD}}, \qquad
\lambda_\ell^{LD} = {q \over p} {\overline{A}_\ell^{LD} \over A_\ell^{LD}}.
\eeq
Note that since we assume that the SD amplitude carries no strong phase, we have $|\lambda_\ell^{SD}|=1$. 

We now adopt one more 
working assumption, that is, we consider only models where: 
\begin{itemize}
\item[$(iii)$]
{\it The long-distance physics is CP conserving.} \\
That is, we only consider cases where $A_\ell^{LD}$ is relatively real,
that is, $\Im\left(\lambda_\ell^{LD}\right)=0$. 
\end{itemize} 
In particular, this assumption implies that we can trust the 
perturbative calculation for the CP-violating amplitude, 
using specific operators described by quarks. 

We are now ready to discuss the CP-odd amplitudes.
Because of assumption $(ii)$ we have $A_1^{\text{CP-odd}} = 0$. Thus we only need to consider the $\ell=0$ CP-odd amplitude. Using Eqs.~(\ref{eq:defacpeo})
and (\ref{eq:acpoddeven}) we write it as
\beq
A_0^\text{CP-odd}= \frac{1}{\sqrt{2}} A_0^{SD}\,(1 + \lambda_0^{SD})\,.
\eeq
Then, using the fact that $|\lambda_0^{SD}|=1$, we get
\beq \label{eq:ftocpamp}
|A_0^\text{CP-odd}|^2 = |A_0^{SD}|^2 \left[1 + \mathrm{Re}(\lambda_0^{SD} ) \right] 
= |A_0^{SD}|^2 \left[1 - \cos \left(2\phi^{SD}_0\right)\right]
		=  2 |A_0^{SD}|^2 \sin^2\phi^{SD}_0\,.
\eeq
where we define
\beq
\phi^{SD}_0 = {1 \over 2} \mathrm{arg}\left(-\lambda_0^{SD} \right).
\eeq
Note that the result is independent of the way we choose to split the
amplitude into long- and short-distance physics as long as the part we
call ``long-distance'' is relatively real. Moreover, we can
subtract from $A^{SD}_0$ any part that is relatively real without
affecting the result. We use this freedom below when we discuss the SM prediction.

We conclude that in any model that satisfies our assumptions, we need
to calculate $|A_0^{SD}|^2$ and $\sin^2\phi^{SD}_0$ in order to make a
prediction for ${\cal B} (K_S \to \mu^+\mu^-)_{\ell=0}$. 

\subsection{SM calculation}
Next we discuss the situation in the SM and remark on more generic
models. The SM short-distance prediction has been discussed in
Ref.~\cite{Isidori:2003ts}. Here we do not present any new arguments,
but instead we review the results in the literature, explicitly
stating the assumptions made, 
and present the results in a basis independent way.

\begin{figure}[t]
 \begin{center}
  \includegraphics[width=0.8\textwidth]{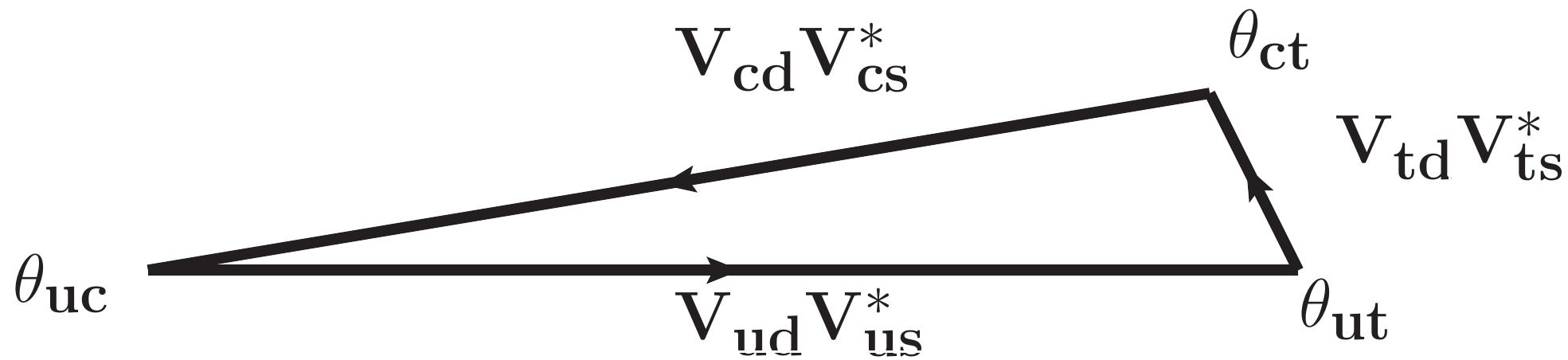}
  \caption{The \lq\lq{}ds\rq\rq{} unitarity triangle, see
    Refs.~\cite{Aleksan:1994if, Lebed:1996ay}. The plot is not to scale.}
\label{fig:triangle}
\end{center}
\end{figure}

In order to discuss the situation in the SM we look at the
\lq\lq{}ds\rq\rq{} unitarity triangle, that we plot 
in Fig.~\ref{fig:triangle}. The angles are given as \cite{Lebed:1996ay}:
\begin{align}
\theta_{ct} &\equiv \mathrm{arg}\left(-\frac{V_{td} V_{ts}^*}{V_{cd} V_{cs}^*}\right)
= \pi - \beta-\beta_s \sim \lambda^0\,, \\
\theta_{ut} &= \mathrm{arg}\left(-\frac{ V_{ud} V_{us}^*}{V_{td} V_{ts}^*} \right)
	    = \beta+\beta_s-\theta_{uc}\sim \lambda^0\,, \\
\theta_{uc}&= \mathrm{arg}\left(-\frac{ V_{cd} V_{cs}^*}{V_{ud} V_{us}^*} \right)\sim \lambda^4\,.	    
\end{align}
In what follows, when we discuss the SM prediction, we make one more
approximation:
\begin{itemize}
\item[$(iv)$] We neglect effects of $\mathcal{O}(\lambda^4)$. In particular, we set
  $\theta_{uc}=0$.
\end{itemize}
With this approximation we then write 
\beq
{q \over p} = -\left(\frac{V_{cd} V_{cs}^*}{V_{cd}^* V_{cs}}\right)
\left[1+\mathcal{O}(\lambda^4)\right] \approx -\left(\frac{V_{cd} V_{cs}^*}{V_{cd}^* V_{cs}}\right)\,.
 \label{eq:qoverp}
\eeq
where in the last step we used $\theta_{uc}=0$. 

We are now ready to show that in the SM the long-distance amplitude is
CP conserving, complying with assumption (iii) above.
The claim is that 
the CKM factors in the long-distance amplitudes are to a good
approximation $V_{us} V_{ud}^*$. The reason is that rescattering
effects, which are what results in the long-distance contributions,
are dominated by tree level decays followed by QCD rescattering. 
The most important one is $K \to
\gamma\gamma$, which is dominated by the $\pi^0$ poles
\cite{GomezDumm:1998gw,Isidori:2003ts}. We thus have 
\beq
\lambda^{LD}_0 =  {q \over p} {\overline{A}^{LD}_0 \over A^{LD}_0} =
-\left(\frac{V_{cd} V_{cs}^*}{V_{cd}^* V_{cs}}\right)\left( 
\frac{V_{ud} V_{us}^*}{V_{ud}^* V_{us}}\right)  \quad \Rightarrow \quad \Im(\lambda^{LD}_0)=0.
\eeq
where in the last step we use 
$\theta_{uc}=0$.  The fact that $\Im( \lambda^{LD}_0) = 0$ implies 
that the long-distance amplitude is CP conserving.

We next discuss working assumption (ii) above, that is, that CP violation enters
only for $\ell=0$. 
Within the SM the short-distance effects are due to the following Hamiltonian
\beq\label{eq:SMeffH}
{\cal H}_{\rm eff} = -\frac{G_F}{\sqrt{2}}\frac{\alpha}{2\pi\sin^2\theta_W} \left[V_{cs}^*V_{cd} Y_{NL}+ V_{ts}^*V_{td}Y(x_t)\right] \left[(\bar s d)_{V-A}(\bar \mu \mu)_{V-A}\right] + h.c.,
\eeq
with $x_t = m_t^2/m_W^2$, and the loop function $Y(x_t)\approx 0.950\pm 0.049$ and 
$Y_{NL}={\cal O}(10^{-4})$~\cite{Buchalla:1995vs, Gorbahn:2006bm}.
Thus the leading SM short-distance physics operator is
\beq
(\bar s d)_{V-A}(\bar \mu \mu)_{V-A} + h.c.\label{eq:v-aop}\,.
\eeq 
This operator contributes only to the $\ell=0$ final state~\cite{Isidori:2003ts}. 
For completeness, we provide a short derivation of this known result in Appendix~\ref{sec:l0-only}.

A few comments are in order: 
\begin{enumerate}
\item 
Scalar operators could also lead to CP violation in the $\ell=1$
amplitude through short-distance effects. However, in the SM, the contribution of
these operators to the rate are suppressed with respect to the
operator in Eq.~\eqref{eq:SMeffH} by a factor of $(m_K/m_W)^2\sim 10^{-5}$~\cite{Hermann:2013kca}, 
and can be safely neglected for the extraction of SM parameters. 
\item
Only the axial-times-axial part of the hadronic times leptonic currents  of Eq.~(\ref{eq:v-aop})
is relevant for $K \to \mu^+ \mu^-$ (see Appendix~\ref{sec:l0-only}).
\end{enumerate}

We conclude that the approximations and assumptions we work under are
valid in the SM up to very small deviations, of order $\lambda^4 \sim
\epsilon_K\sim 10^{-3}$.
Thus, within the SM, the only source of a CP violating phase is the weak effective Hamiltonian given in Eq.~\eqref{eq:SMeffH}. 
Moreover, any extension of the SM in which the leptonic operator
remains vectorial rather than a scalar would satisfy our set of
assumptions. For example also models with right-handed currents fall under this category. Thus, within the SM and any such extension it is straightforward to extract a prediction for ${\cal B} (K_S \to \mu^+\mu^-)_{\ell =0}$ purely from short-distance physics.

We are now ready to discuss the SM prediction for ${\cal B} (K_S \to \mu^+\mu^-)_{\ell =
0}$. We recover the result, given in
Ref.~\cite{Isidori:2003ts}, using phase
convention independent expressions (see Appendix~\ref{sec:appSM}).
We first redefine $A_0^{SD}$ by subtracting the
charm contribution, which is relatively real under the approximation
$\theta_{uc}=0$. Then we can write
\beq\label{eq:lambdaSD}
\lambda_0^{SD} =  {q \over p} {\overline{A}_0^{SD} \over A_0^{SD}} =
-\left(\frac{V_{cd} V_{cs}^*}{V_{cd}^* V_{cs}}\right)\left( 
\frac{ V_{td}^* V_{ts}}{V_{td} V_{ts}^*}\right) = -e^{-2i\theta_{ct}}
\quad \Rightarrow \quad \sin^2\phi_0^{SD}= \sin^2\theta_{ct}.
\eeq

The calculation of  $|A_0^{SD}|^2$ and the phase space integral is
reviewed in Appendix~\ref{sec:appSM}. The result is given in Eq.~(\ref{eq:rate-SM}):
\beq\label{eq:A0CPoddSM}
{\cal B} (K_S \to \mu^+\mu^-)_{\ell = 0} &=& {\frac{\beta_\mu \,\tau_S}{16\pi m_K}}\left|
\frac{G_F}{\sqrt{2}} \frac{2\alpha_{em}}{ \pi\sin^2\theta_W} m_K
m_{\mu} \times  Y(x_t)\, \times  \, f_K \times\, V_{ts}V_{td} \sin\theta_{ct}
 \right|^2 \\ \nonumber
&\approx& 1.64 \cdot 10^{-13} \,\times \, \left|\frac{V_{ts}V_{td} \sin\theta_{ct}}{(A^2\lambda^5\bar\eta)_\text{best fit}}\right|^2 \, ,
\eeq
where we use
\beq
(A^2\lambda^5\bar\eta)_\text{best fit}= 1.33\cdot 10^{-4}.
\eeq

Eq.~(\ref{eq:A0CPoddSM}) is very precise. There are a few sources of
uncertainties that enter here. They are all under control:
\begin{enumerate}
\item
The only hadronic parameter
is the kaon decay constant, which
is well known from charged kaon decays. 
Isospin breaking effects can also be incorporated in lattice QCD if
needed~\cite{Aoki:2019cca}, reducing the ultimate hadronic uncertainties
below the $1\%$ level.
\item
We have neglected subleading terms, that is, we
neglected the term proportional to $Y_{NL}\sim 10^{-4}$ from
Eq.~(\ref{eq:SMeffH}), as well as CPV effects of order $\epsilon_K$.
\item
Parametric errors, including the dependence of
the loop function $Y(x_t)$ on $m_t/m_W$, are small, as the errors on the
top and $W$ masses are below the $1\%$ level.
\item
Only leading order results in the loop expansion are used. 
Higher order terms are expected to be suppressed by a loop factor,
which is of order $1\%$. If needed, higher orders 
in the loop function can be incorporated in
order to reduce this uncertainty.
\item
We only consider the leading SM operator, which is vectorial. At higher order scalar operators are also present, but these effects are suppressed by 
$\mathcal{O}(m_K^2/m_W^2)$ \cite{Hermann:2013kca}.
\end{enumerate}
We conclude that a measurement of ${\cal B} (K_S \to \mu^+\mu^-)_{\ell = 0}$ 
would be a very
clean independent measurement of the following combination of CKM
elements 
\beq
\left|V_{ts} V_{td} \,\sin \theta_{ct} \right|= \left|V_{ts} V_{td} \,\sin (\beta+\beta_s) \right|\approx A^2 \lambda^5 \bar\eta\,,
\eeq
which coincides with Eq.~(\ref{eq:ImVtsVtd}).

A similar analysis can be done in any model that satisfies the
assumptions we have made. In particular, these results hold in any model that generates the same operator as in the SM. In such a model the prediction would be amended by replacing the SM values for the CKM parameters and the
loop function with the respective values in the model under
consideration.

We end this section with two remarks
\begin{enumerate}
\item
There are models where we can have a significant contribution to the
CP-odd amplitude from scalar operators~\cite{Chobanova:2017rkj}, in
which case our assumption $(ii)$ is not satisfied.
\item
In addition to our quantity of interest, ${\cal
  B}(K_S\to\mu^+\mu^-)_{\ell=0}$, under the same set of assumptions it
is also possible to calculate the short-distance contribution to
$A_0^\text{CP-even}$, that is, $A(K_L\to \mu^+\mu^-)_{\ell=0}^{SD}$. 
Then, assuming given values for the CKM parameters, the measurement of the interference terms is also a measurement of the long-distance amplitude $A(K_L\to \mu^+\mu^-)_{\ell=0}^{LD}$, and in particular of its unknown sign~\cite{DAmbrosio:2017klp}.
\end{enumerate}

\section{Experimental considerations}

We now turn to discuss the feasibility of the extraction of
${\cal B}(K_S\to\mu^+\mu^-)_{\ell=0}$. As is apparent from Eq.~(\ref{eq:magicBR}) we need to experimentally extract $C_{Int.}$ and $C_L$. Of these, $C_L$ has
already been measured, and we can expect that in the future it will 
be measured with even higher precision. The question is how
well can $C_{Int.}$ be extracted.

\begin{figure}[t]
 \begin{center}
  \includegraphics[width=1.0\textwidth]{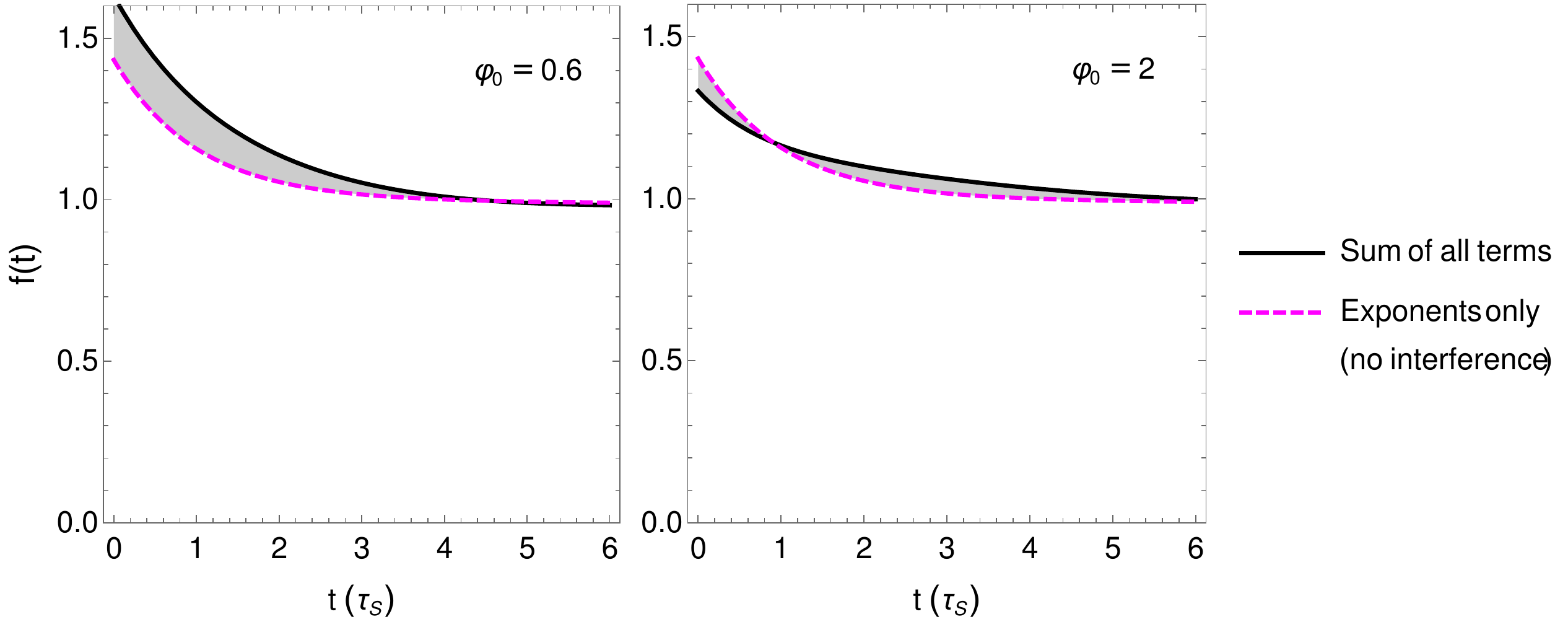}
  \caption{The expected approximate time dependence within the SM, using the coefficients of Eq.~(\ref{eq:Csnum}), for two values of 
$\varphi_0= \arg({A_0^\text{CP-odd}}^*A_0^\text{CP-even})$. 
The difference between the dashed magenta curve and the solid black one is a measure of interference
effects.  }
\label{fig:sketch}
\end{center}
\end{figure}

Below we estimate the number of kaons that is needed to perform
the measurement assuming the
SM. For that
we need the values in the SM of the relevant amplitudes.
While the method we discuss 
does not require any estimation of the amplitudes, we use these
estimates 
to illustrate the expected magnitude of the interference terms, and to
estimate the needed statistics to perform the measurements.
Of the three amplitudes, $|A_0^\text{CP-odd}|$  can
be calculated perturbatively, 
$|A_0^\text{CP-even}|$ can be extracted directly from the measured
value of $\mathcal{B}(K_L\to\mu^+\mu^-)$, and $|A_1^\text{CP-even}|$
can only be estimated a priori by relying on non-perturbative calculations from the literature, that suffer from large hadronic uncertainties. We provide the details of these estimations in Appendix~\ref{sec:appSM}. They result in the following values for the experimental parameters:
\beq\label{eq:Csnum}
	(C^{K^0}_L)_{\rm SM} &=&  |A_0^\text{CP-even}|^2 \equiv 1, \\ \nonumber
	(C^{K^0}_S)_{\rm SM} &=&  |A_0^\text{CP-odd}|^2 + \beta_\mu^2|A_1^\text{CP-even}|^2 \approx 0.43, \\ \nonumber
	(C^{K^0}_{Int.})_{\rm SM} &=& |A_0^\text{CP-even}||A_0^\text{CP-odd}| \approx 0.12, 
\eeq
where we have used a normalization such that the coefficient $(C^{K^0}_L)_{\rm SM}$ is set to be unity. Using these estimates, we plot the time dependence of the rate
in Fig.~\ref{fig:sketch}, for two values of the unknown phase, 
$\varphi_0= \arg({A_0^\text{CP-odd}}^*A_0^\text{CP-even})$. For illustration, we also plot the time dependence excluding the interference terms (see caption). 
We use the range $t \lesssim 6 \tau_S$ as for larger times the beam is
almost a pure $K_L$ beam.
The relative magnitude of the interference terms is apparent in the
difference between the two plotted curves. We find the relative
integrated effect to be of order $3\%$ to $6\%$, depending on the value of $\varphi_0$.

Based on the above, we can roughly estimate the number of required kaons. We have 
$\mathcal{B}(K_L\rightarrow \mu^+\mu^-) = (6.84\pm 0.11)\cdot 10^{-9}$~\cite{Zyla:2020zbs},
and only about $1\%$ of the $K_L$ particles decay inside our region of interest,
$t \lesssim 6 \tau_S$. Since the coefficients in Eq.~(\ref{eq:Csnum})
are not very small, we can use this to estimate that the number of
useful events is roughly a fraction of $10^{-10}$ out of the kaons. Thus, for
example, in order to get ${\cal O}(1000)$ events in the interesting region we need
${\cal O}(10^{13})$ $K^0$ particles to start with. 
We do not expect this preliminary estimate to be strongly affected by backgrounds or reconstruction efficiencies. 

Experimentally, it is not easy to produce a pure neutral kaon beam. 
Experiments currently running enjoy a very high luminosity of
kaons of order $10^{14}$
kaons a year (see Ref.~\cite{Kleimenova:2019pcu} for NA62,
Ref.~\cite{Ahn:2020opg} for KOTO, and Ref.~\cite{Junior:2018odx} for LHCb).
However, these kaons are either charged (NA62), or to a good
approximation a pure $K_L$ (KOTO), or come with an
almost equal mix of $K^0$ and $\overline K{}^0$ (LHCb). 

Thus, for the purpose of the analysis we are considering, 
we need to turn to a mixed beam or a regenerated beam. As discussed in
Appendix~\ref{sec:tDandr}, in the case of 
a mixed beam with non-zero production
asymmetry, the sensitivity to the interference terms
is diluted by a factor of $D$. The use of matter effects, for
example in the case of a $K_L$ beam going through a regenerator,
introduces suppression that is proportional to the regeneration parameter, $r$. Thus, the number of kaons
that are needed in these cases compared to the pure kaon beam, 
are larger by roughly $1/D$ or $1/r$ as we need to
overcome these suppression effects.

Several approaches that could be useful in acquiring 
the needed sensitivity to the interference terms appear in the literature:
\begin{enumerate}
\item
There are cases with QCD production where both $\kzero$ and $\kbar$ are
produced, but there is an asymmetry, that is $D \ne 0$. One example is the ``high intensity $K_S$-run'' at the NA48 experiment, which reported $10^{10}$ $K_S$ decays with $D\sim 0.3$~\cite{Mazzucato:2000qv}. 
\item
Regeneration in $K_L$ beams \cite{Good:1961ik, Bohm:1968ewa,
  Kleinknecht:1973ny, Angelopoulos:1997gw}. Numerically, typical values for 
$r$ range 
from ${\cal O}(10^{-2})$ to a few times $10^{-1}$, depending on the 
material and on the relevant kaon momentum. 
\item
The use of a charge exchange target in order to generate pure $K^0$ beams from $K^+$ beams~\cite{Siegel:1984br,Mehlhop:1968nka}.
\item
Post-selection using tagging in high energy production, for example, by looking at the charge of the pion in $K^*$ decays, or by tagging $\Lambda^0$ and $K^-$ in $pp\to
K^0K^- X$ and $pp\to K^0 \Lambda^0 X$ decays~\cite{DAmbrosio:2017klp}.
\end{enumerate}

We do not discuss these options in any detail. The high yields of
currently running experiments is encouraging in terms of the
ability of future endeavors to reach the desired sensitivity, should
some of
these methods be implemented. Clearly, a detailed study of the
experimental requirements is needed in order to arrive at a reliable estimate for the expected sensitivity.

We close this section with a remark about the time dependence. A measurement of the full time dependence would result in the best sensitivity. However,
in principle, a measurement of the integral over four different time intervals suffices to get the needed information. In practice, $C_L$ is already known, $C_S$ can
be extracted from a beam with $D=0$, and then we would need two time
intervals using a beam with $D \ne 0$ or $r\ne 0$.

\section{Conclusion and Outlook}
We have demonstrated how, under well-motivated approximations and assumptions, it
is possible to cleanly test the SM using a measurement of the time-dependent decay rate of $K\to\mu^+\mu^-$. 
A necessary ingredient is sensitivity to the interference between the
$K_L$ and $K_S$ amplitudes, as can be seen from Eq.~(\ref{eq:magicBR}),
which is our main result.
The relevant SM parameter of interest is 
\beq
\left|V_{ts}V_{td}\,\sin (\beta+\beta_s)\right|,
\eeq
which is exactly the CKM parameter combination that appears in $K_L\to
\pi^0\nu\bar{\nu}$ . Thus, our proposal is to use
$K\to\mu^+\mu^-$ as an additional independent measurement of the same
SM quantity.

As we discuss in detail, the point to emphasize is that the extraction
is theoretically very clean. There are several assumptions that were
made in setting up the method, as well as in the calculation within the
SM. All of these are valid within the SM to a few per-mill, giving a total
uncertainty below the $1\%$ level. This is comparable to the best probing
methods for the angle $\beta$ and related quantities, that is, 
the CP asymmetries in $B \to \psi K_S$ and the decay rate of  
$K_L \to \pi^0 \nu \bar\nu$.
The assumptions we rely on are additionally respected by any extension
of the SM in which the relevant leptonic current is vectorial. 

The approach we discuss can in principle be extended to other decay
modes. Most promising are the decays $K \to \pi e^+ e^-$ and $K \to \pi \mu^+
\mu^-$. The generalization is not trivial as these decays involve
more partial waves beyond $\ell=0,1$. We plan to discuss these modes in a future publication.

Our very preliminary estimates indicate that these measurements can be carried out
in next generation kaon experiments. This is encouraging, and 
more detailed feasibility studies are called for.

\begin{acknowledgments}
We thank G.~D'Ambrosio, T.~Kitahara, and Y.~Nir for useful comments on the manuscript.
The work of AD is partially supported by the Israeli council for higher education postdoctoral fellowship for women in science. The work of YG is supported in part by the NSF grant PHY1316222.
S.S.~is supported by a Stephen Hawking Fellowship from UKRI under reference EP/T01623X/1. 
\end{acknowledgments}


\appendix

\section{Extracting ${\cal B} (K_S \to \mu^+\mu^-)_{\ell =0}$ without
  a pure kaon beam} \label{sec:tDandr}

In the main text, we demonstrated how we can determine
${\cal B} (K_S \to \mu^+\mu^-)_{\ell =0}$ for the case of a pure $K^0$ beam in empty
space. Here we present a discussion on two other cases
which are more related to realistic experimental situations. The first
case is when we have a beam with unequal initial number of $K^0$ and
$\kbar$. The second case is when we have a pure $K_L$ beam going
via a regenerator before the kaons decay. In both cases, it is possible to extract the branching ratio ${\cal B} (K_S \to \mu^+\mu^-)_{\ell =0}$ cleanly as in Eq.~\eqref{eq:magicBR}, with the addition of a dilution factor as in Eq.~\eqref{eq:magicBRgen}.

\subsection{A mixed beam of $K^0$ and $\kbar$}

Consider a beam which initially consists of an incoherent mixture of kaons and anti-kaons. We define the production asymmetry,
\beq 
D = \frac{N_{K^0}-N_{\kbar}}{N_{K^0}+N_{\kbar}} .
\eeq
such that the fractions of $K^0$ and $\kbar$ particles are given
respectively by
\beq 
\frac{N_{K^0}}{N_{K^0}+N_{\kbar}} = \frac{1+D}{2}, \qquad
\frac{N_{\kbar}}{N_{K^0}+N_{\kbar}} = \frac{1-D}{2}.
\eeq
Note that $D=1$ corresponds to a pure $K^0$ beam, while $D=-1$ corresponds to a pure $\kbar$ beam.

The decay rate to a final state $f$ is given by the incoherent sum
\begin{align}
{\D \Gamma \over \D t} &= \frac{1+D}{2}\left({\D \Gamma_{K^0} \over \D t}\right) + \frac{1-D}{2}\left({\D \Gamma_{\kbar} \over \D t}\right) ,
\end{align}
such that  its form is given by Eq.~\eqref{eq:Cdef}
with the following coefficients:  
\beq
C_L &=& |A_0^\text{CP-even}|^2 , \\ \nonumber
C_S &=& |A_0^\text{CP-odd}|^2 + \beta_{\mu}^2 |A_1^\text{CP-even}|^2,
\\ \nonumber
C_{cos} &=& D\,|A_0^\text{CP-odd*}A_0^\text{CP-even}|\cos\varphi_0, \\ \nonumber
C_{sin} &=& D\,|A_0^\text{CP-odd*}A_0^\text{CP-even}|\sin\varphi_0.
\eeq
It is then straightforward to extract our parameter of interest.  For
$D \neq 0$ we obtain
\beq
|A_0^\text{CP-odd}|^2 =	{\cal D}_F\frac{{C_{cos}}^2 +
  {C_{sin}}^2}{C_L}, \qquad {\cal D}_F = \frac{1}{D^2} \,.
\eeq
In terms of the branching ratios we have
\beq \label{eq:magicBRD}
{\cal B}(K_S \to \mu^+\mu^-)_{\ell=0} = {\cal D}_F \times
{\cal B}(K_L \to \mu^+\mu^-) \times {\tau_S \over \tau_L} \times \left({C_{int} \over C_L}\right)^2,
\eeq
We learn that the beam asymmetry serves as a dilution factor compared to the case of a pure
$K^0$ or $\kbar$ beam. Note that if $D=0$ (which means that the beam
is an equal admixture of $K^0$ and $\kbar$), one cannot use the beam to measure ${\cal B}(K_S \to \mu^+\mu^-)_{\ell=0}$.

We close with a remark regarding the LHCb search for the $K_S$ rate \cite{Aaij:2020sbt}.
To a very good approximation at LHCb we have $D=0$. In that case the
interference terms cancel and we are left with just $C_L$
and $C_S$. Thus, without any further analysis to tag the flavor of the
kaon, LHCb is working on extracting the $C_S$ term that includes the
decay to both the $\ell=0$ and $\ell=1$ states. 

\subsection{$K_L$ propagating through a slab of matter}

When kaons travel through matter, the time dependence of the kaon
wave function is modified via the inclusion of the momentum dependent
regeneration parameter~\cite{Good:1961ik, Bohm:1968ewa, Kleinknecht:1973ny, Fetscher:1996fa, Angelopoulos:1997gw, Quinn:2000jy}.

We define
\beq 
r e^{i \alpha} =-\frac{\pi N}{m}\left(\frac{\Delta f}{\Delta\lambda } \right),\eeq
where $r$ and $\alpha$ are real, and
\beq
\Delta f\equiv f-\bar f, \qquad 
\Delta\lambda \equiv \Delta m-\frac{i}{2}\Delta\Gamma.
\eeq
Here, $f$ $(\bar f)$ is the difference of forward scattering
amplitudes for kaons (anti-kaons), and $N$ is the density of
scattering centers in the regenerator. Note that $r$ and $\alpha$ are physical and can be determined from
experiment.

Let us consider a pure $K_L$ beam, which is produced by letting the
$K_S$ (and interferences) terms decay away. Then we put 
a regenerator of length $L$ in the path of the beam. Let $t_L$ be the
time taken by the kaon to travel through the regenerator. We
define $t=0$ to be the time the kaon
emerges from the regenerator. We then study the time dependence of
the kaon wave function at later times. For simplicity, in the following
we present the result to
leading order in $r$.

The normalized decay rate is given by Eq.~\eqref{eq:Cdef}, with the coefficients:  
\beq
C_L &=& |A_0 ^{\text{CP-even}}|^2,\nonumber \\
C_S &=& 0,\nonumber  \\
C_{sin} &=& - r\,|A_0 ^{\text{CP-even}} A_0 ^{\text{CP-odd}}| \left( \sin(\alpha-\varphi_0)-e^{t_L \Delta \Gamma/2} \sin (\alpha -\varphi_0 + \Delta m t_L)\right),\nonumber \\
C_{cos} &=&  r\,|A_0 ^{\text{CP-even}} A_0 ^{\text{CP-odd}}| \left(\cos(\alpha-\varphi_0)-e^{t_L \Delta \Gamma/2} \cos (\alpha -\varphi_0 + \Delta m t_L) \right).
\eeq
We can check that for $t_L$ = 0 
(which means that the regenerator thickness is negligible) or $r=0$
(the regenerator material is just vacuum), the interference terms
vanish as it should.

Using the above, we find the dilution factor ${\cal D}_F$, for $t_L \neq 0$ and $r\neq 0$
to be
\beq 
{\cal D}_F =  \frac{1}{2 r^2}\left(\frac{\cosh(\Delta \Gamma
    t_L/2)-\sinh(\Delta \Gamma t_L/2)}{\cosh(\Delta \Gamma
    t_L/2)-\cos(\Delta m t_L)}\right). 
\eeq
We learn that the dilution parameter depends on both $r$ and $t_L$. 
The extraction of the rate is given by Eq.~\eqref{eq:magicBRD}.

We close with two remarks
\begin{enumerate}
\item
As we already emphasized, the interference terms are the key to the
extraction of ${\cal B}(K_S \to \mu^+\mu^-)_{\ell=0}$. Having $D\neq0$ or $r\neq
0$ are some of the ways of obtaining interference terms in the time
dependence of the kaon beam. 
\item
More generally one may also have combinations with both non-zero $D$
and $r$, as well as a general initial state. 
The calculation is straightforward, though tedious and does
not provide much further insight, and so we do not show it here. 
\end{enumerate}

\section{SM Calculations}
\label{sec:appSM}

In the following we first derive the SM prediction for ${\cal B}(K_S
\to \mu^+\mu^-)_{\ell=0}$, and then derive approximate numerical
estimates 
for the experimental parameters within the SM.

\subsection{SM calculation}

Using the standard formula for two body decays~\cite{Zyla:2020zbs}, as
well as the results of
Eqs.~(\ref{eq:lambdaSD}) and (\ref{eq:ftocpamp}), we write
\beq \label{eq:genBR}
{\cal B}(K_S\to (\mu^+\mu^-)_{\ell=0}) =  \frac{\beta_\mu \tau_S}{16\pi m_K} \times 2 \sum|{\cal M}^{SD}(K_S\to
(\mu^+\mu^-)_{\ell=0})|^2 \times \sin^2\theta_{ct}, 
\eeq
where the sum is over the outgoing spin, as usual.
Note that ${\cal M}$ is proportional to $A$, defined in
Eq.~(\ref{eq:Af-Afbar}) but it uses the
standard normalization that is used when making calculations.

We write the matrix element for the short-distance contribution as 
\beq
{\cal M}^{SD} = g_{\rm SM} \bra{\mu\bar{\mu}}{\cal O}_\ell\ket{0}\times \bra{0}{\cal O}_H\ket{K},
\eeq
where
\beq
{\cal O}_\ell = (\bar{\mu}_L\gamma^\rho \mu_L), \qquad 
\bra{0}{\cal O}_H\ket{K} \equiv  -i\, p_K^\rho \, f_K.
\eeq
For the kaon decay constant we employ here the convention
\begin{align}
\bra{0} \bar{s} \gamma_{\mu} \gamma_5 d\ket{K^0(p)} &= i p_{\mu} f_{K^0} \,.
\end{align}
The coupling, $g_{\rm SM}$, can be read from Eq.~(\ref{eq:SMeffH}) (note that $(\bar{\mu}\mu)_{V-A} = 2 (\bar{\mu}_L\gamma^\rho \mu_L)$)
\beq
g_{\rm SM} =  
 -\frac{G_F}{\sqrt{2}}\frac{\alpha_{em}}{\pi\sin^2\theta_W} 
\left[V_{cs}^*V_{cd} Y_{NL}+ V_{ts}^*V_{td}Y(x_t)\right].
\eeq
Since under our assumption of $\theta_{uc}=0$ the part proportional to
$V_{cs}^*V_{cd}$ is relatively real, we can further define
\beq
\tilde g_{\rm SM}=  
 -\frac{G_F}{\sqrt{2}}\frac{\alpha_{em}}{\pi\sin^2\theta_W} 
V_{ts}^*V_{td}Y(x_t).
\eeq
As long as what we are after is the CP violating decay rate, we can use
$\tilde g_{\rm SM}$.

Squaring the amplitude and summing over spins, we find
\beq \label{eq:SMcal}
\sum|{\cal M}^{SD}_{K\to\mu\mu}|^2&=& \Big[-p_K^\rho p_K^\sigma
f_K^2\Big] \,|g_{\rm SM}|^2\,{\rm Tr}\Big[\bar u(k_1) \gamma_\rho P_L v(k_2)\bar v(k_2)P_L \gamma_\sigma u(k_1)\Big] \\ \nonumber
	&=& - |g_{\rm SM}|^2 \, f_K^2  p_K^\rho p_K^\sigma {\rm Tr}\Big[\gamma_\rho P_L(\slashed k_2-m_\mu)P_L\gamma_\sigma (\slashed k_1+m_\mu)\Big] \\ \nonumber
	&=&  |g_{\rm SM}|^2 \, f_K^2 m_\mu^2  p_K^\rho p_K^\sigma {\rm Tr}\Big[\gamma_\rho \frac{1}{2}(1-\gamma_5)\gamma_\sigma\Big]\\ \nonumber
	&=& 2|g_{\rm SM}|^2 f_K^2 m_\mu^2  m_K^2.
\eeq
Using Eq.~(\ref{eq:genBR})
we find 
\beq \label{eq:rate-SM}
{\cal B}(K_S\to (\mu^+\mu^-)_{\ell=0})&= & \frac{\beta_\mu \tau_S}{16\pi m_K}
 4 |\tilde g_{\rm SM}|^2 f_K^2 m_\mu^2 m_K^2 \sin^2\theta_{ct} \\
&=& 
{\frac{\beta_\mu \,\tau_S}{16\pi m_K}}\left|
\frac{G_F}{\sqrt{2}} \frac{2\alpha_{em}}{ \pi\sin^2\theta_W} m_K
m_{\mu} \times  Y(x_t)\, \times  \, f_K \times\,
V_{ts}V_{td} \sin\theta_{ct} \right|^2, \nonumber
\eeq
in agreement with Eqs. (37) and (39) of Ref.~\cite{Isidori:2003ts}.

We next get numerical estimates. We use the lattice QCD result~\cite{Aoki:2019cca} for the hadronic parameter, assuming isospin symmetry:
\begin{align}
f_K &= 155.7\pm 0.3\,{\rm MeV}\,.
\end{align}
We use the following values for the measured parameters~\cite{Zyla:2020zbs},
\beq
	&\,& m_K = 497.61\,{\rm MeV}, \qquad\qquad \quad\quad\,\,\,  m_\mu=105.658\,{\rm MeV}, \\ \nonumber
	&\,& G_F = 1.166378\times 10^{-5}\,{\rm GeV^{-2}}, \,\quad\;\, \alpha_{em}=1/129, \\ \nonumber
	&\,& \sin^2\theta_W= 0.23, \qquad\qquad\quad\qquad \quad\,\, Y(x_t)=0.95,\\ \nonumber
	&\,& \tau_{L} = 5.116\times 10^{-8}\,{\rm s},\qquad\quad\qquad \quad\,\tau_{S} = 8.95\times 10^{-11}\,{\rm s},
\eeq
and for the CKM values we use 
\beq
	\left|V_{ts}V_{td}\,\sin\theta_{ct}\right| = A^2\lambda^5\bar\eta,\qquad  \text{with}\quad  A=0.79,\,\lambda=0.2265,\, \bar\eta=0.357,
\eeq
to arrive at the prediction 
\beq\label{eq:CP-oddNum}
{\cal B}(K_S\to (\mu^+\mu^-)_{\ell=0}) \approx 1.64 \cdot 10^{-13} \,\times \, \left|\frac{V_{ts}V_{td} \sin\theta_{ct}}{(A^2\lambda^5\bar\eta)_\text{best fit}}\right|^2 \, ,
\eeq
with
\beq
 	(A^2\lambda^5\bar\eta)_\text{best fit} = 1.33\cdot 10^{-4}.
\eeq

\subsection{SM approximate values for the experimental parameters}
In order to estimate the magnitude of the effect we are after and to illustrate the expected time dependence, we require approximate values for the remaining two branching ratios within the SM.
First, we use the measured branching ratio,
\beq
	\mathcal{B}(K_L\to \mu^+\mu^-)_{\rm exp.} = \mathcal{B}(K_L\to (\mu^+\mu^-)_{\ell=0})_{\rm exp.} = (6.84\pm 0.11)\cdot 10^{-9},
\eeq
which sets the value of the parameter $C_L$.

The remaining branching ratio, ${\cal B}(K_S\to \mu^+\mu^-)_{\ell=1}$,
can only be estimated a priori by relying on non-perturbative
calculations from the literature that suffer from large hadronic
uncertainties. Nonetheless, we use these results to get an estimate
for its magnitude. Below we use the prediction for the long-distance contribution,~\cite{DAmbrosio:2017klp}   
\beq
\mathcal{B}(K_S\to \mu^+\mu^-)_{\rm SM}^{LD}= {\cal B}(K_S\to \mu^+\mu^-)_{\ell=1} &\approx & 4.99\cdot 10^{-12}.
\eeq
Note that while we quote results to three significant digits, the
theoretical uncertainties are much larger.
Altogether we have
\beq
	{\cal B}(K_S\to \mu^+\mu^-)_{\ell=0} &\approx & 1.64 \cdot 10^{-13}, \\ \nonumber
	{\cal B}(K_L\to \mu^+\mu^-)_{\ell=0}&\approx & 6.84\cdot 10^{-9}, \\ \nonumber
	{\cal B}(K_S\to \mu^+\mu^-)_{\ell=1} &\approx & 4.99\cdot 10^{-12}.
\eeq
The first is the result of the calculation from the SM effective
Hamiltonian, the second is the experimental measured value, and the
third uses the non-perturbative estimation together with the calculated SM short-distance contribution.

For illustration of the time dependence, we choose to normalize the $C$ coefficients such that $C_L=1$.
The numerical values for the coefficients, as defined in
Eqs.~({\ref{eq:Cdef}) and ({\ref{eq:defCInt}), for the case of a pure $K^0$ or $\overline K{}^0$ beam, are then given by:
\beq
	(C_L)_{\rm SM} & \equiv & 1, \\ \nonumber
	(C_S)_{\rm SM} &=&  \frac{\tau_L}{\tau_S}\frac{{\cal B}(K_S\to\mu^+\mu^-)_{\ell=0}+ {\cal B}(K_S\to\mu^+\mu^-)_{\ell=1}}{{\cal B}(K_L\to\mu^+\mu^-)_{\ell=0}}\approx 0.43, \\ \nonumber
	(C_{Int.})_{\rm SM} &=& \sqrt{\frac{\tau_L \,{\cal B}(K_S\to\mu^+\mu^-)_{\ell=0}}{\tau_S\,{\cal B}(K_L\to\mu^+\mu^-)_{\ell=0}}}\approx 0.12\,.
\eeq

There is one more experimental parameter, the phase $\varphi_0$. It
is related to the strong phase and we do not provide any estimate for it.

\section{The short-distance operator}
\label{sec:l0-only}

For completeness, we explain below the well-known results that 
the short-distance SM amplitude cannot generate an $\ell=1$ state, and
that only the axial parts of both the hadronic and leptonic currents contribute in two-body pseudo-scalar decays.

Our starting point is the factorization of the matrix element
\beq \label{eq:facto-M}
{\cal M} = \langle \mu^+\mu^- | O_L^{\mu} O_{H \mu}| K \rangle = 
\langle \mu^+\mu^- | O_L^{\mu}| 0 \rangle\times  \langle 0| O_{H\mu} | K \rangle ,
\eeq
where
\beq
O_H^{\mu} = (\bar{s} d)_{V-A}, \qquad O_L^{\mu} = (\bar{\mu} \mu)_{V-A} .
\eeq
The leading breaking of this factorization is from the photon loop, and
thus it is suppressed by roughly ${\cal O} (\alpha_{EM}/4 \pi)\sim 10^{-3}$. 

Considering the leptonic part is sufficient to explain why short-distance physics does not contribute to the $K \to (\mu^+ \mu^-)_{\ell =1}$ amplitude.
For two spinors $\psi$ and $\chi$, we recall the transformation of the $V-A$ operator under CPT~\cite{Branco:1999fs}:
\begin{align}
\Theta\bar{\psi}\gamma^{\mu}(1 - \gamma^5) \chi  \Theta^{\dagger} &= 
 -\bar{\chi} \gamma^{\mu}(1 - \gamma^5) \psi\,, \label{eq:Cptransf}
\end{align}
where $\Theta = {\cal CPT}$ is the CPT operator. This implies
\beq
\Theta O_L^\mu \Theta^\dagger = -O_L^\mu.
\eeq
Using 
\beq
\Theta |\mu^+\mu^-\rangle_\ell= (-1)^{\ell+1}
|\mu^+\mu^-\rangle_\ell, \qquad
\Theta |0\rangle= |0\rangle,
\eeq
we conclude
\beq
\langle (\mu^+\mu^-)_\ell | O_L^\mu| 0 \rangle= 
\langle (\mu^+\mu^-)_\ell| \Theta\Theta^{\dagger}
O_L^\mu \Theta\Theta^{\dagger}| 0 \rangle = (-1)^\ell \langle (\mu^+\mu^-)_\ell | O_L^\mu| 0 \rangle.
\eeq
From the above we see that ${\cal M}$, defined in Eq.~(\ref{eq:facto-M}), vanishes when $\ell$ is odd.

As for the axial part of the hadronic current, the argument is the same as the well-known one
for charged pion decay, that we recall below. Consider
\beq
\langle 0 |V^\mu-A^\mu|K(p_K)\rangle.
\eeq
The kaon is a pseudo-scalar and the vacuum is parity-even. Thus, the matrix element of $V^\mu$ must transform as a pseudovector, and the matrix element of $A^\mu$ must transform as a vector. The only available physical observable is the kaon momentum, $p_K$, which is a vector. It is impossible to construct a product of any number of $p_K^\mu$ that transforms like as a pseudovector. We conclude that
\beq
\langle 0 | V^\mu|K\rangle =0.
\eeq

In order to see that also for the leptonic current only the axial part is relevant, we write the matrix element in the following form, leaving the vector and axial-vector components of the leptonic operator general:
\beq
 {\cal M} \sim p_K^\alpha \,\,\bar u(k_2)\gamma_\alpha (B+A\gamma^5)v(k_1)
\eeq
Then,
\beq\label{eq:trace}
	\sum|{\cal M}|^2 &\propto & {\rm Tr}\Big[(\slashed k_2 + m_\mu)\slashed p_K (B+A\gamma^5)(\slashed k_1 -m_\mu)(B^*+A^*\gamma^5)\slashed p_K\Big] \\ \nonumber
	& = & 4\big(|B|^2-|A|^2\big)\Big[2(k_1\cdot p_K)(k_2\cdot p_K)-m_K^2(k_1\cdot k_2)\Big]\, - \, 4\big(|B|^2+|A|^2\big)m_\mu^2 m_K^2
\eeq
Using two-body kinematics, we have
\beq
	(k_1\cdot p_K) &=& (k_2\cdot p_K) = \frac{1}{2}m_K^2, \\ \nonumber
	(k_1\cdot k_2)&=& \frac{1}{2}m_K^2-m_\mu^2.
\eeq
Plugging this in to Eq.~(\ref{eq:trace}), the $|B|^2$ terms drop out and we are left only with the terms proportional to $|A|^2$,
\beq
	\sum|{\cal M}|^2 \propto |A|^2 m_\mu^2 m_K^2,
\eeq
 i.e., only the axial-vector part of the operator is relevant.

\bibliography{DGGS}
\bibliographystyle{apsrev4-1}

\end{document}